\documentclass[twocolumn,pre]{revtex4-1}
\usepackage{color}
\usepackage{graphics,graphicx,epsfig,wrapfig}
\usepackage{epsf,epstopdf}
\usepackage{amssymb,amsfonts,amsmath}
\usepackage{enumitem}
\usepackage{setspace}

\usepackage{ifthen}

\newcommand\mydots{\hbox to 0.8em{.\hss.\hss.}}
\newcommand{\beqn}{\begin{eqnarray}}
\newcommand{\eeqn}{\end{eqnarray}}
\newcommand{\beq}{\begin{equation}}
\newcommand{\eeq}{\end{equation}}

\newcommand{\AW}[1]{{\color{black}#1}}
\newcommand{\ff}{1}

\newcommand{\mytitle}{Precise tracking of vaccine-responding T-cell clones reveals convergent and personalized response in identical twins }
\newcommand{\myauthors}{Mikhail V. Pogorelyy$^{1,2}$, Anastasia A. Minervina$^{1}$, Maximilian Puelma Touzel$^{3}$, \\
Anastasiia L. Sycheva$^{1}$, Ekaterina A. Komech$^{1,2}$, Elena I. Kovalenko$^{1}$, Galina G. Karganova$^{4}$,\\
 Evgeniy S. Egorov$^{1,2}$, Alexander Yu. Komkov$^{1,5}$,
Dmitriy M. Chudakov$^{1,2,6,7}$, Ilgar Z. Mamedov$^{2,1}$,\\
Thierry Mora*$^{8}$, Aleksandra M. Walczak*$^{3}$, Yuri B. Lebedev*$^{1,9}$}

\makeatletter
\def\@seccntformat#1{%
  \expandafter\ifx\csname c@#1\endcsname\c@section\else
  \csname the#1\endcsname\quad
  \fi}
\makeatother

\begin{document}
\title{\mytitle}
\author{\myauthors}
\affiliation{~\\
\normalsize{$^{1}$ Shemyakin-Ovchinnikov Institute of Bioorganic
  Chemistry,}
\normalsize{Moscow, Russia}\\
\normalsize{$^{2}$Pirogov Russian National Research Medical University, Moscow, Russia}\\
\normalsize{$^{3}$ Laboratoire de physique th\'eorique,}
\normalsize{CNRS, Sorbonne Universit\'e, and \'Ecole normale sup\'erieure (PSL), Paris, France}\\
\normalsize{$^{4}$Chumakov Institute of poliomyelitis and viral encephalitides, Moscow, Russia}\\
\normalsize{$^{5}$Dmitry Rogachev National Medical Research Center of Pediatric Hematology, Oncology and Immunology, Moscow, Russia}\\
\normalsize{$^{6}$Center for Data-Intensive Biomedicine and Biotechnology, Skoltech, Moscow, Russia}\\
\normalsize{$^{7}$Masaryk University, Central European Institute of Technology, Brno, Czech Republic}\\
\normalsize{$^{8}$ Laboratoire de physique statistique,}
\normalsize{CNRS, Sorbonne Universit\'e, Universit\'e Paris-Diderot, and \'Ecole normale sup\'erieure (PSL), Paris,
  France}\\
 \normalsize{$^{9}$Moscow State University, Moscow, Russia}\\
\normalsize{\rm *These authors contributed equally.}\\
}

\begin{abstract} 
T-cell receptor (TCR) repertoire data contain information about infections that could be used in disease diagnostics and vaccine development, but extracting that information remains a major challenge. Here we developed a statistical framework to detect TCR clone proliferation and contraction from longitudinal repertoire data. We applied this framework to data from three pairs of identical twins {\color{black} immunized with} the yellow fever vaccine. We identified 500-1500 responding TCRs in each donor and validated them using three independent assays. While the responding TCRs were mostly private, albeit with higher overlap between twins, they could be well predicted using a classifier based on sequence similarity. Our method can also be applied to samples obtained post-infection, making it suitable for systematic discovery of new infection-specific TCRs in the clinic.
\end{abstract}

\maketitle

\section*{Introduction}

The extremely diverse repertoire of T-cell receptor sequences allows the immune system to develop a specific response to almost any possible pathogen. In recent years huge progress has been made in the deep profiling of T-cell receptor repertoires by high throughput sequencing (HTS), allowing for the identification of millions of TCR sequences in a single experiment \cite{Benichou2012}. TCR sequences and phenotypes and the relative abundances of the corresponding T-cell clones encode both the history of previous infections and the protection against yet unseen pathogens. However, despite recent large-scale efforts \cite{Dash2017, Glanville2017}, it is still impossible to predict systematically which antigen is recognized by a TCR with a given sequence.
The most reliable method to identify antigen-specific TCR --- MHC-multimer staining \cite{Davis2011} --- is restricted by the choice of HLA alleles (which are the most polymorphic genes in the human population \cite{Robinson2013}), and by the knowledge of immunodominant peptides. HLA-independent methods for quantitative monitoring of challenge-specific TCRs \textit{in vivo} are still lacking.

Here we developed a methodology for identifying responding TCR$\beta$ clonotypes 
from time-dependent repertoire-sequencing data and applied it to yellow fever immunization: a classical model of acute viral infection in humans. The yellow fever vaccine (YFV strain 17D) is one of the most efficient and safe vaccines ever made \cite{Monath2015}.  Because YFV 17D is a live attenuated virus, vaccination leads to viremia and very intense T-cell and humoral responses \cite{Miller2008}. 
Tracking of activated T-cell subsets such as CD8+CD38+HLA-DR+ \cite{Miller2008} and fluorescent MHC-tetramer staining \cite{Akondy2009} make it possible to quantitatively describe the kinetics of the T-cell response to primary YFV 17D-immunization. The response peaks around day 14 after immunization, when activated T-cells responding to vaccination occupy 2-13\% of the CD8+ subpopulation \cite{Miller2008,Akondy2009,Akondy2015} and 3-4\% of the CD4+ subpopulation \cite{Blom2013,Kohler2012,Kongsgaard2017}. Several immunodominant peptides were identified and the corresponding pMHC-multimers have made it possible to track YFV-specific T-cells years and even decades after immunization \cite{Akondy2009, Kongsgaard2017, FuertesMarraco2015}. The only repertoire sequencing study of yellow fever immunization available 
to our knowledge \cite{DeWitt2015} reports thousands of CD8+ T-cell clones expanding after yellow fever immunization, and preferential recruitment of highly expanded CD8+ T-cell clones to the memory subpopulation. However, the clonal structure of the T-cell immune response, how personalized this response is, and what is the impact of genetic factors on the response still remain poorly understood. Studying monozygous twins allows us  {\color{black}to quantify }
 the impact of genetic factors using a small cohort of donors \cite{Zvyagin2014,Rubelt2016,Qi2016}.

In this study, we identified nearly five thousand YFV 17D-responding T-cell receptors in three pairs of identical twin donors, using HTS profiling of T-cell repertoires at different timepoints analyzed with advanced statistical modeling. We validated the yellow fever specificity of expanded clones using three independent functional tests. The detailed analysis of the TCR$\beta$ sequences of the expanded clones showed both a highly personalized response and high sequence similarity across individuals, especially between twins. This convergence allowed us to develop a supervised classifier that predicted YFV-reactive TCR sequences with high specificity. Remarkably, the dynamics of clonal contraction in the month after the peak response specifically predicts the YFV-reactive clonotypes. Thus, our methodology can be used during the post-infection period in the clinic to identify TCR clonotypes that recognized and responded to an acute infection of interest, even without prior knowledge of donor MHC-alleles or pathogen epitopes. 

\section*{Results}

 \begin{figure*}
\noindent\includegraphics[width=\ff\linewidth]{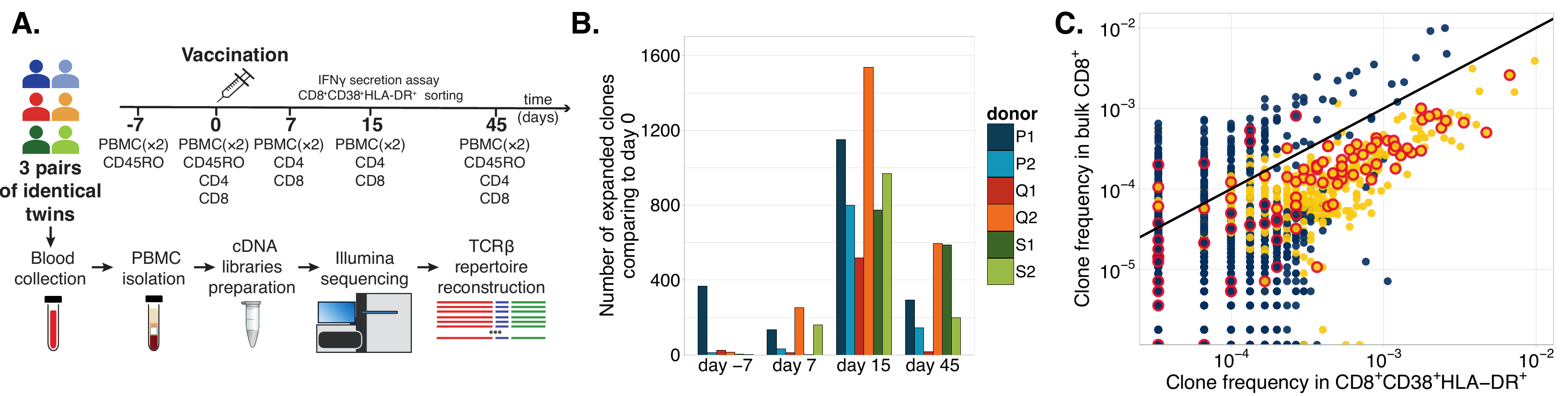}
\caption{{\bf A. Yellow fever vaccination study design.}  Top: the experimental timeline with the list of samples collected at each timepoint. Unsorted PBMC samples were collected in two biological replicates at each timepoint. Bottom: method overview. Peripheral blood samples were subjected to PBMC isolation, synthesis of TCR$\beta$ cDNA libraries, Illumina sequencing, and reconstruction of TCR$\beta$ repertoires.
{\bf B. The number of significantly expanded clonotypes in comparison to day 0.} The number peaks at day 15 for all donors. The comparison to day -7 corresponds to a contraction reflecting the normal dynamics of a healthy repertoire in absence of vaccination.
{\bf C. Activated CD8+CD38+HLA-DR+ subpopulation is enriched with clonotypes expanded between days 0 and 15.}  The relative abundance of a clonotype in the CD8+CD38+HLA-DR+ activated subpopulation (x-axis) is plotted against the relative abundance in the bulk CD8+ population isolated at the same timepoint (y-axis). Yellow dots indicate clonotypes that strongly expanded between day 0 and day 15. Black line shows identity. Red circles indicate that the clonotype was found in the A02-NS4b$_{214-222}$-dextramer-positive fraction 2 years later.
}
\label{fig1}
\end{figure*}

\subsection*{Detection of significantly expanded TCR$\beta$ following YFV 17D-vaccination}

Blood samples were obtained for three pairs of identical twin volunteers aged 20-23. We collected peripheral blood samples on 5 different timepoints (2 before and 3 after immunization) with the live attenuated YFV-17D vaccine (see Fig. 1A). On each timepoint, we collected two biological replicates --- two independent tubes of blood --- to isolate bulk PBMC, and another tube of blood to isolate CD4+ and CD8+ T-cell subpopulations. Additional portions of blood were used for other subpopulations isolation (CD45RO+) and functional tests on several timepoints (see Methods for details). From each sample, cDNA libraries of TCR$\beta$ chains were prepared as previously described \cite{Pogorelyy2017}, and sequenced on the Illumina HiSeq platform.

We developed a Bayesian statistical framework that identifies T-cell clonotypes both significantly ($p < 0.05$) and strongly (fold change $> 32$) expanded between different timepoints compared to the expected variability between replicates (see Methods). We reproduced the results with the edgeR package \cite{Robinson2009}: a widely used method to analyze differential gene expression in RNAseq experiments (see Methods). In Fig 1B we show the number of clonotypes identified as expanded with respect to day 0. In all donors we observe many more expanded clonotypes between day 0 and day 15 than for any other pair of timepoints, despite variations between donors. In the following we simply call {\em expanded clonotypes} TCR$\beta$ sequences that significantly increased in fraction between days 0 and 15. Clonotypes with a significantly higher frequency on the pre-vaccination timepoint relative to the vaccination timepoint (day 0 vs day -7, so actually corresponding to a contraction) were relatively few for all donors except for P1, who we speculate was undergoing another transient immune response.

Most expanded clones were not detected before immunization, and often not even on day 7, due to their low frequency. We report expansion rates of up to 2000-3000 fold in 7 days, although this estimate is only a lower bound due to lack of detection prior to day 15 (see SI Fig. S4). 

\subsection*{The majority of expanded TCR$\beta$ are YFV 17D-specific}

We hypothesized that most of the expanded clonotypes proliferated specifically in response to the YFV-17D vaccine. To check this hypothesis, three independent functional tests with subsequent TCR$\beta$ repertoire sequencing were performed on donor S1: {\em (i)} interferon (IFN) gamma secretion assay; {\em (ii)} fluorescent sorting of the activated CD8+CD38+HLA-DR+ subset, which was reported to be largely YFV 17D-specific at the peak of the response \cite{Miller2008}; {\em (iii)} staining with a MHC-dextramer loaded with an immunodominant epitope. For {\em (i)}, whole blood was incubated with the YFV-17D vaccine and IFN-gamma producing cells were isolated using magnetic beads (see Methods). For {\em (i)} and {\em (ii)}, we considered the clonotype validated if it was enriched in the  IFN-gamma positive {\em (i)} or CD8+CD38+HLA-DR+  {\em (ii)} fractions, by comparison to the bulk PBMC or CD8+ populations respectively. Enrichment was determined by a one-sided exact Fisher test (see Methods). We found that out of 774 clonotypes expanded in donor S1 (see Fig 1B.), 331 were enriched in the CD8+CD38+HLA-DR+ fraction (see Fig. 1C), and 64 were enriched in the IFN-gamma secretion assay. For {\em (iii)} we used the HLA-A*02 dextramer loaded with the immunodominant YFV epitope NS4b$_{214-222}$ (LLWNGPMAV) on a sample collected 2 years after immunization. Dextramer-positive cells were isolated and the TCR$\beta$ repertoire was sequenced. We found 68 expanded clonotypes in the A02-NS4b$_{214-222}$ dextramer positive fraction. We  used this data to  estimate the fraction of response to the immunodominant epitope A02-NS4b$_{214-222}$ in this donor. We found that at least 22\% of the CD8+ response to YFV vaccination (identified by our statistical model) consisted of NS4b$_{214-222}$-specific clonotypes on day 15 after immunization. We also sequenced unsorted PBMC for the 2 year timepoint of this donor and found that the fraction of the repertoire occupied by the yellow fever responding clones largely declined 2 years after immunization: from 2.1\% (on day 45) to 0.07\%, but still exceeded prevaccination levels by nearly two orders of magnitude.

Overall, the total number of clonotypes validated by any of the three methods was 395 out of 774. Notably, the remaining unvalidated clonotypes had significantly lower average frequencies than the validated ones on day 15 (t-test p-value$<10^{-16}$, SI Fig. S5). We performed a randomization control to check the specificity of our validation approach: 774 clonotypes were randomly subsampled from the repertoire on day 15, with the probability of being sampled taken to be proportional to the clonotype frequency. We then tested this random subset for enrichment in the same way we did for the expanded clonotypes. On average, over 1000 such randomizations, we found $29.2(\pm 0.15)$ random clonotypes enriched in the CD8+CD38+HLA-DR+ subset (compared to 331 in the actual expanded clones), $8.3(\pm 0.1)$ clonotypes enriched in the IFN-gamma fraction (vs. 64 among expanded clones) and $6.4(\pm 0.1)$ clonotypes enriched in the A02-NS4b$_{214-222}$ dextramer positive fraction (vs. 68 among expanded clones). Note that this control gives a conservative estimate of validation specificity, because the expanded clones (true positives) are still likely to be sampled in this randomization due to their high frequency on day 15.

To further validate the expanded clones for YFV-17D specificity, we used previously published TCR$\beta$ sequences from the VDJdb database \cite{Shugay2017} (\url{https://vdjdb.cdr3.net}, see SI data) that lists sequences specific for the the NS4b$_{214-222}$ YFV-17D immunodominant epitope. We checked if these published TCR$\beta$ sequences were present in our donors, and whether they were expanded after vaccination. We screened the CD8+ TCR$\beta$ repertoires of our donors at different timepoints for A02-NS4b$_{214-222}$ specific sequences, as well as for sequences specific for an unrelated epitope as a control: the cytomegalovirus (CMV) immunodominant HLA-A*02-restricted pp65$_{495-503}$ peptide (NLVPMVATV). {\color{black} All of our donors are HLA-A*02-positive (complete HLA genotypes can be found in SI Table S1).}
We found that both the number of hits and their cumulative frequency (SI Fig. S1) were significantly higher for published A02-NS4b$_{214-222}$-specific sequences on day 15 than on day 0 (one-sided t-test p=0.019). By contrast, in the control no significant difference was found (one-sided t-test p=0.27). Then we checked if published A02-NS4b$_{214-222}$-specific sequences could be found in sets of expanded clonotypes identified in each donor by our model (SI Table S2). We found multiple exact matches for published A02-NS4b$_{214-222}$-specific sequences and no exact matches for CMV-specific A02-pp65$_{495-503}$ clonotypes.

In summary, the expanded clonotypes were validated for their YFV specificity, and were also consistent with published YFV-specific TCRs.

\begin{figure}
\noindent\includegraphics[width=\linewidth]{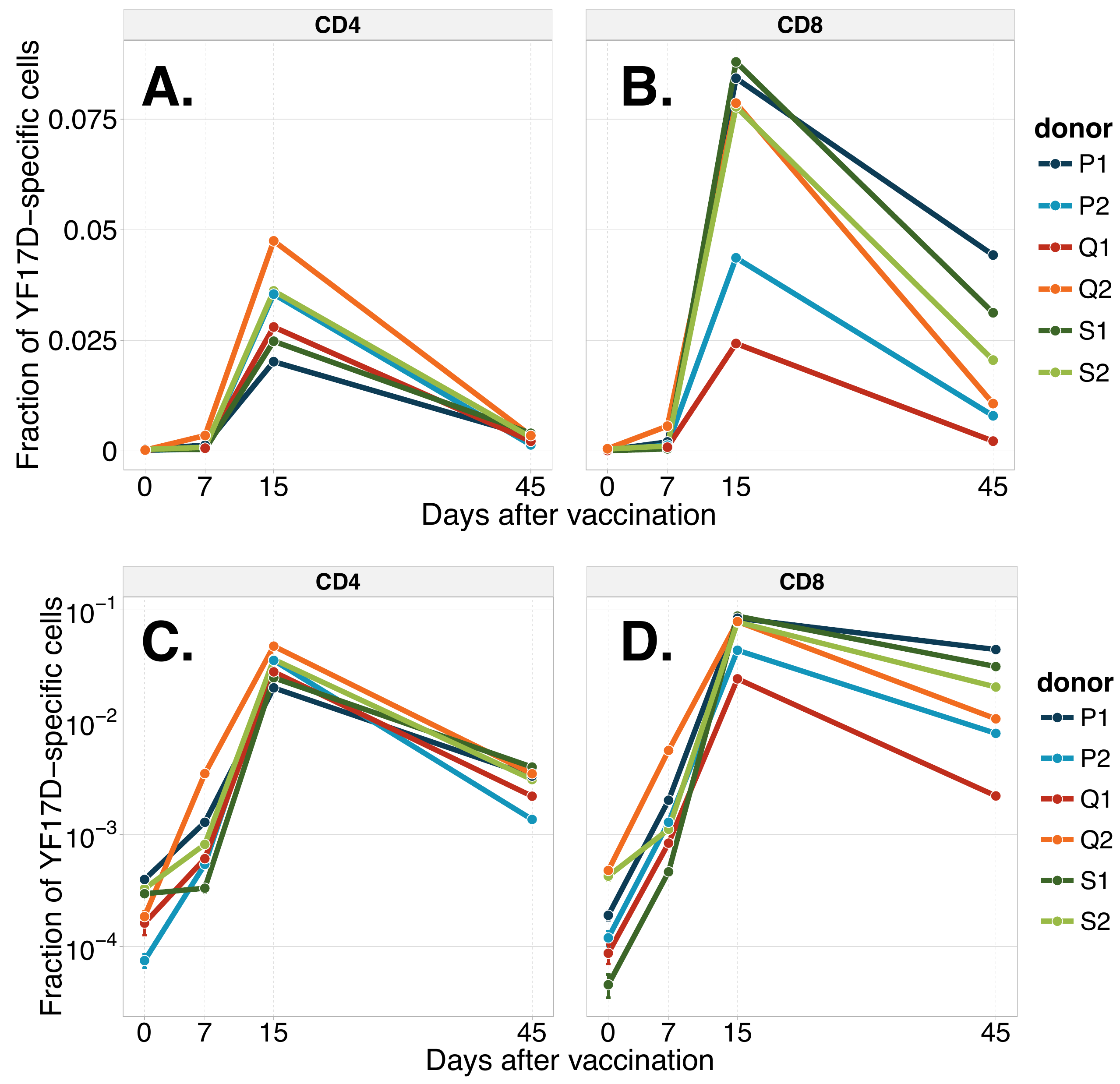}
\caption{{\bf Dynamics of YFV-specific T-cells in the CD4+ and CD8+ compartments.}  The total fraction of ({\bf A}) CD4+ and ({\bf B}) CD8+ repertoires occupied by clonotypes significantly expanded from day 0 to day 15 for different timepoints. CD4+ and CD8+  T-cell subpopulations show similar dynamics, although the CD8+ response degrades more slowly. Error bars are smaller than one line width. Panels {\bf C.} and {\bf D.} show the same data on a logarithmic scale. 
}
\label{fig2}
\end{figure}

\subsection*{The CD8+ T-cell response is sustained longer than the CD4+ response}

To track the fractions of the CD4+ and CD8+ repertoires involved in the response over time, we calculated the cumulative frequency of expanded clonotypes in the CD4+ and CD8+ compartments at each timepoint. We found similar dynamics in all donors, with both CD8+ and CD4+ responses peaking on day 15. In all donors the CD4+ response fell off more quickly than the CD8+ response from day 15 to day 45, with a mean $12.8 (\pm 3)$ fold decrease for CD4+, versus $5.4 (\pm 1.4)$ for CD8+ ($p=0.023$, paired one-sided t-test). We checked the significance of this difference in decay for each donor separately using our statistical framework (see Methods). Briefly, for each clonotype we redrew a random number of counts on days 15 and 45, with the same variability as inferred from replicate experiments to mimic experimental noise. Repeating this process many times gave a p-value corresponding to the fraction of numerical experiments where the ordering between CD4+ and CD8+ decay was opposite to the actual observation. CD4+ and CD8+ fold decreases significantly differed ($p<0.0001$) for 5 donors out of 6, with the exception of donor Q1 ($p=0.15$) who was an outlier in the CD8+ response strength and the number of significantly expanded clones. The responding clonotypes at the peak of the response occupied up to 8\% of the repertoire (cumulative frequency) for CD8+ and  up to 5\% for CD4+ T-cells. Almost all these clonotypes were undetected before immunization.

The previous analysis described the dynamics of the CD4+ and CD8+ responses separately. What is the relative importance and diversity of the CD4+ and CD8+ responses within the overall T-cell population? To answer this question, we associated expanded clonotypes from the unsorted PBMC sample to either the CD4+ or CD8+ subsets in the following way: we labeled an expanded TCR$\beta$ clonotype from the unsorted PBMC sample CD4+ if it had a larger concentration in the CD4+ subpopulation sequenced than in the CD8+ subpopulation, and vice versa. This procedure gave unambiguous {\em in silico} phenotypes for each expanded clonotype sequence from the PBMC at day 15: the relative abundance of sequences in the CD4+ versus CD8+ compartments was strongly bimodal, with two peaks close to 100\% CD4+ and 100\% CD8+ (see SI Fig. S2). This analysis revealed that both CD4+ and CD8+ clones strongly expanded in response to the vaccination, with no strong preference for CD4+ or CD8+ expanded clonotypes in terms of the cumulative frequency or diversity (see SI Fig. S3). We next asked if the presence of expanded clonotypes in the memory (CD45RO+) compartment 45 days after immunization depends on their CD4+ or CD8+ phenotype. We detected on average 49\% ($\pm 7$ \%) of CD4+ expanded clonotypes in memory repertoires 45 days after immunization, versus only 21\% ($\pm 3$ \%) of CD8+ expanded clonotypes. This may be explained by different levels of CD45RO+ expression in CD4+ and CD8+ memory T-cells reactive to yellow fever. 

\begin{figure*}
\noindent\includegraphics[width=0.8\linewidth]{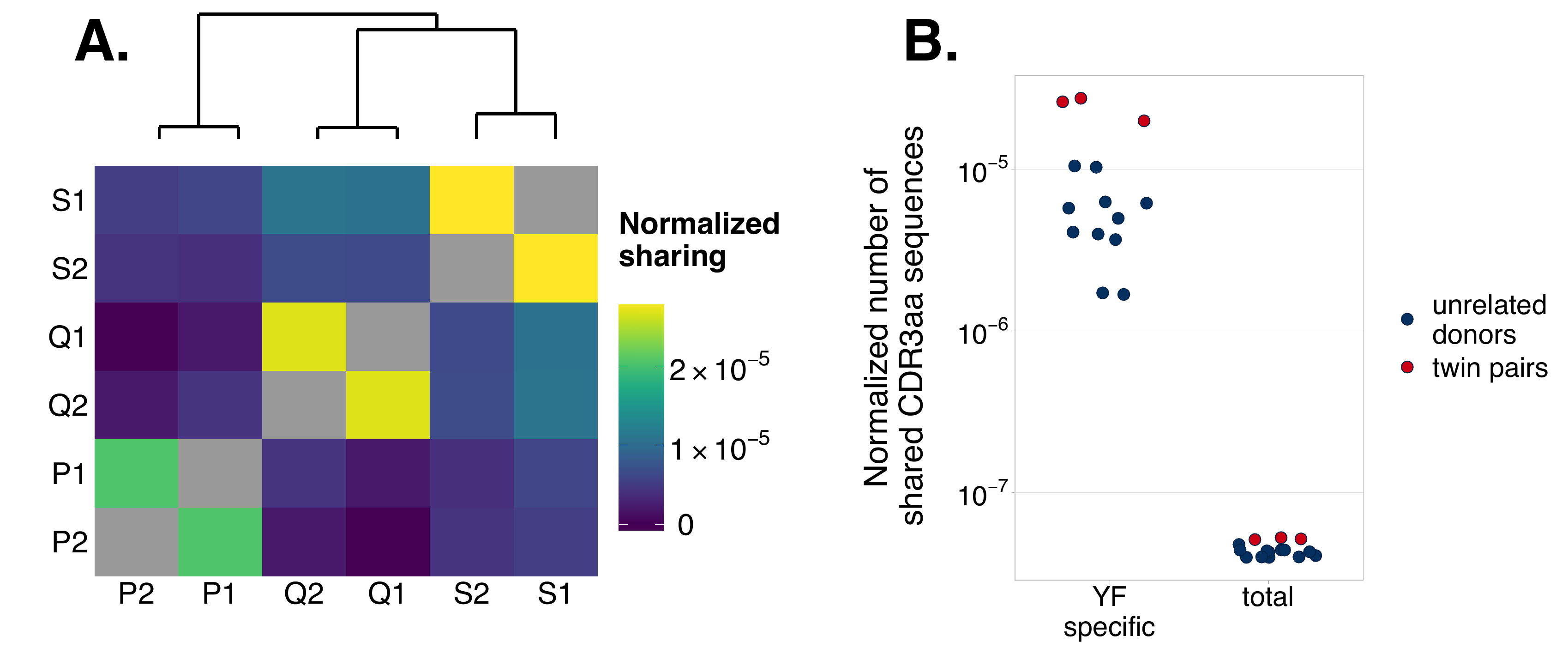}
\caption{{\bf The normalized number of identical yellow fever vaccine (YFV-17D) specific TCR$\beta$ amino acid sequences between different donor pairs.  (A) }  The heatmap shows the number of the identical TCR$\beta$ clonotypes between two sets of clones significantly expanded from day 0 to day 15, divided by the product of sizes of these sets. All twin pairs show higher numbers of identical expanded clonotypes than unrelated individuals and are clustered together by hierarchical clustering (dendrogram on top). {\bf (B)} The normalized sharing of YFV 17D-specific TCR$\beta$ clonotypes (left) is much higher than the normalized sharing in the whole repertoire (right). Sharing in twins (red circles) exceeds sharing in unrelated individuals (dark blue circles) in all cases.
}
\label{fig3}
\end{figure*}

\subsection*{TCR response is highly personalized even among twins}

Our method not only reconstructs the dynamics of immune response, but also enables the analysis of TCR sequences of the responding T-cell clones. For each donor we found 500-1500 YFV-17D reactive clonotypes expanded between day 0 and day 15.  We compared the pairwise sharing of responding amino acid TCR$\beta$ sequences between different individuals. We found more overlap in twins than in non-related individuals (see Fig. 3A). However absolute numbers of identical expanded TCR$\beta$ clonotypes were low even in twin pairs (up to 20 out of 1503 amino acid sequences), indicating that each individual developed an almost unique response. We also observed that the same nucleotide variants encoded some of the amino acid TCR$\beta$ sequences shared between the twins (see SI Table S4). This identity of nucleotide variants was more frequent in twins than expected from convergent recombination alone ($p<0.021$, see Methods). 

Overall, shared clonotypes (across at least two donors) accounted for a small fraction of the response in each donor. Only 2.5\%-4.4\% of unique responding TCR$\beta$ sequences were public (present in more than one donor). These clonotypes correspond to 2.7\%-34\% of YF-specific cells (see SI Table S3). Despite the low numbers of shared clonotypes, normalized sharing of YFV-17D reactive amino acid clonotype sequences was more than two orders of magnitude higher than in the total repertoire both for twins and unrelated individuals (Fig. 3B). This may result from convergent selection of the same TCR variants recognizing the same epitopes in different donors. 

\begin{figure*}
\noindent\includegraphics[width=0.8\linewidth]{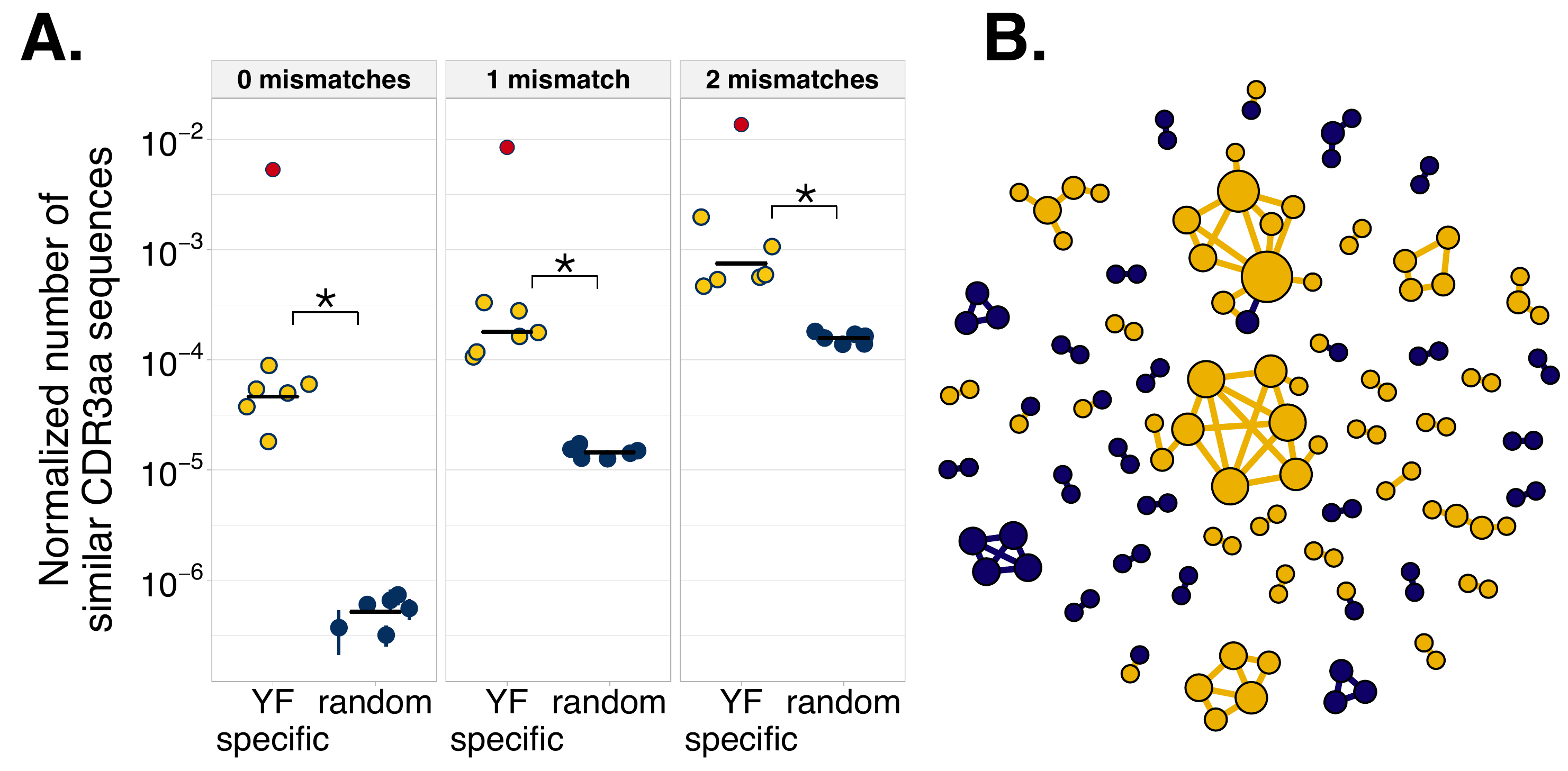}
\caption{{\bf Convergence of amino acid sequences inside the YFV-reactive TCR repertoire. (A) }  The number of pairs of similar clones (exact same CDR3 amino acid sequence - left panel, up to 1 mismatch - middle panel, up to 2 mismatches - right panel) normalized by the number of possible pairings in each individual (see Methods). The number of similar clones in the data (yellow dots) is larger than the number of similar clones in randomly drawn samples (dark blue dots) of the same size (one sided paired t-test, $p=0.022,\ 0.015,\ 0.008$ for 0 , 1, and 2 mismatches respectively). As a reference, the red dots show an example of a restricted and specific repertoire for the yellow fever virus immunodominant epitope NS4b$_{214-222}$ (data from VDJdb \cite{Shugay2017}, see SI data). {\bf (B)} 1000 most abundant TCRs from donor S1 at day 15. Each vertex corresponds to a TCR$\beta$ amino acid sequence; edges connect clonotypes differing by 2 or fewer amino acids in their CDR3. Yellow clonotypes indicate expanded TCRs, while blue clonotypes were present before immunization at similar frequencies as on day 15. The vast majority of edges (95 out of 103) are formed between TCRs of the same status (expanded or not expanded).
}
\label{fig4}
\end{figure*}

\subsection*{TCR sequence analysis reveals a mixture of convergent and private response}

It was previously shown that in many cases TCRs recognizing the same antigens have restricted sequence diversity \cite{Dash2017,Miles2011}. To analyze the sequence diversity of responding TCRs, we performed a pairwise comparison of all expanded TCR$\beta$ sequences within each donor at the amino-acid level. In each individual, we identified many more pairs of expanded TCRs with the same or highly similar CDR3$\beta$ amino acid sequences (up to 0, 1, or 2 aa mismatch) on day 15 than in a random subset of equal size (see Fig. 4A). Interestingly, our expanded TCRs were still more diverse than published TCRs selected for their specificity to a single immunodominant YFV epitope NS4b$_{214-222}$ (red dots, data from VDJdb \cite{Shugay2017}, see SI data), suggesting that the response is directed against multiple epitopes in each donor.
Consistent with the idea of multiple specificities, the expanded TCRs form multiple dense clusters of highly similar clones (Fig. 4B, yellow circles). By contrast, pre-vaccination abundant TCRs form fewer, sparser, and smaller clusters (blue circles).

\begin{figure*}
\noindent\includegraphics[width=0.96\linewidth]{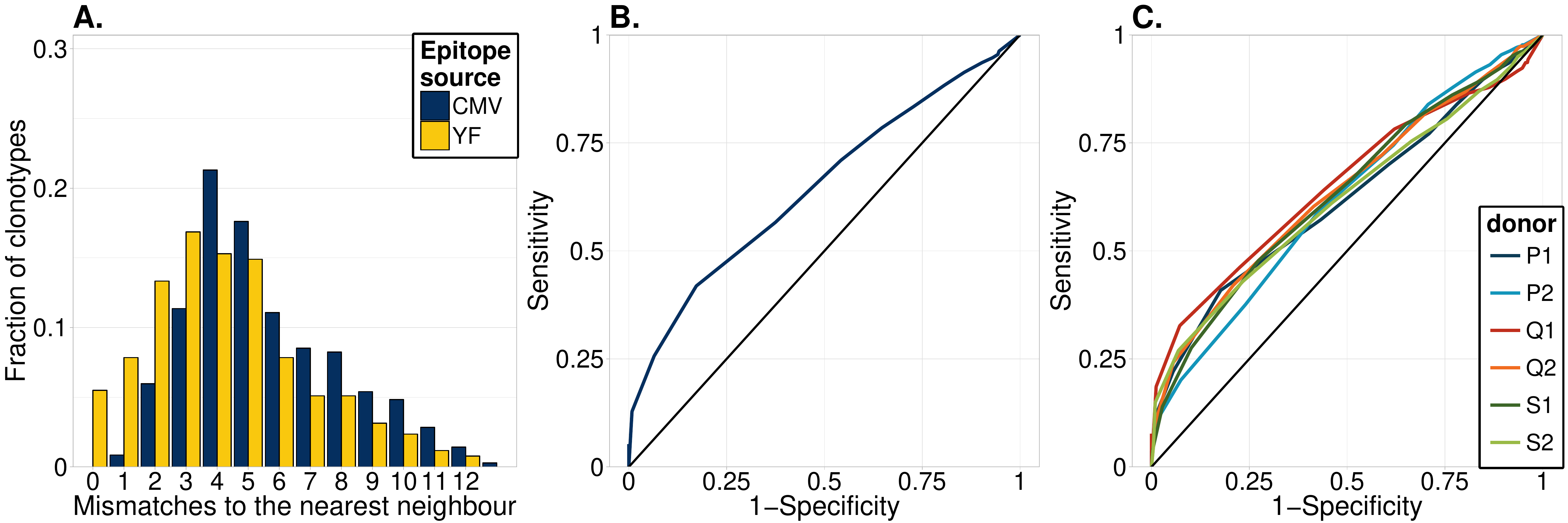}
\caption{{\bf Predicting the YFV specificity of TCR sequences. (A)} Histogram of the minimal Hamming distance of expanded TCR amino acid sequences identified in this study to published YFV-specific sequences (YFV immunodominant epitope NS4b$_{214-222}$, yellow) and cytomegalovirus (CMV)-specific sequences (CMV immunodominant epitope pp65$_{495-503}$, blue). A much larger fraction of our expanded TCRs were similar (0, 1, or 2 mismatches) to YFV-specific sequences than to CMV-specific sequences. {\bf (B)} Receiver Operating Characteristic (ROC) curve for the classification of published NS4b$_{214-222}$-specific sequences using the minimal distance to the expanded TCRs identified in this study. {\bf (C)} ROC curves for ``leave-one-out'' classifier validation. The top 1000 most abundant TCRs of each individual were classified as YFV-reactive or non YFV-reactive according to their minimal distance to TCRs expanded in the other five individuals.}
\label{fig5}
\end{figure*}

\begin{figure}
\noindent\includegraphics[width=0.8\linewidth]{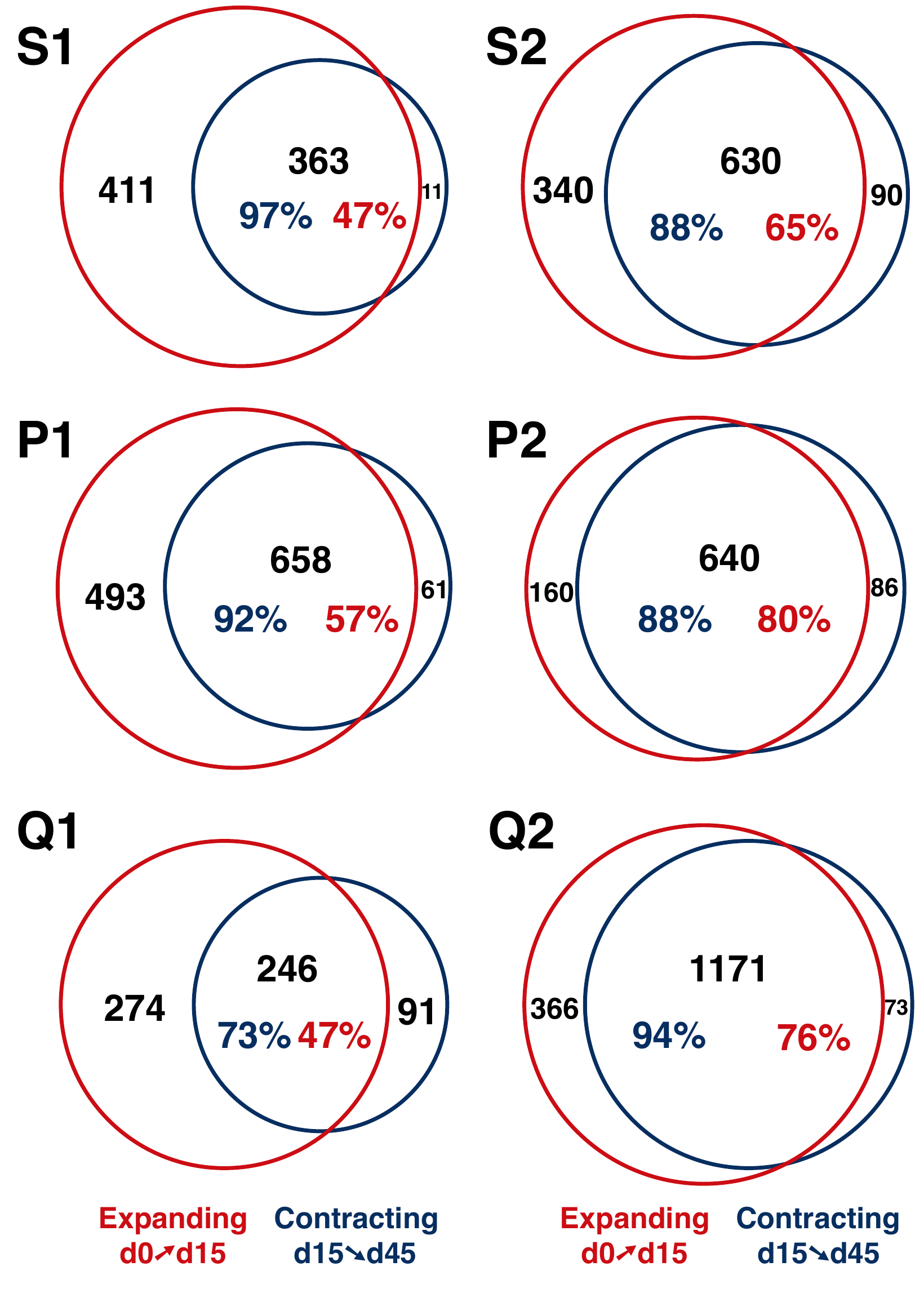}
\caption{{\bf Clonal contraction can be used to identify YFV-17D reactive clonotypes with high specificity and good sensitivity.} Venn diagrams show overlap between the subset of TCR clones significantly expanded from day 0 to day 15 (red), and the subset of TCR clones significantly contracted from day 15 to day 45 (blue). All donors show large overlaps between the contracting and expanding subsets.
}
\label{fig6}
\end{figure}

\subsection*{Sequence score based on distance to expanded TCR predicts YFV-17D specificity}

The similarity of sequences in the expanded repertoire makes it possible to build a simple classifier to identify novel YFV 17D-specific clonotypes. For each TCR of interest we defined a YFV specificity score as the Hamming distance to the closest amino acid sequence neighbour among the expanded clonotypes from all 6 donors. To test how informative this score is about YF specificity, we first applied it to published TCRs (from VDJdb \cite{Shugay2017}, SI data) specific to the NS4b$_{214-222}$ epitope from YFV-17D and to the pp65$_{495-503}$ epitope from CMV as a negative control. We found that published YF-specific clonotypes were much closer to our set of expanded clones than CMV-specific clonotypes (Fig. 5A), with some exact matches for YF-specific but none for CMV-specific sequences (SI Table S2).
Accordingly, using the score to discriminate YF-specific from CMV-specific published sequences yields good specificity and sensitivity (Fig. 5B).

We then asked if the score could be used to identify reactive clonotypes in the repertoire of an individual after immunization. To test this capability, we used a leave-one-out approach. Information about expanded clonotypes in five individuals was used to build a score as described earlier. The score was then used to predict expanded clonotypes among the 1000 most abundant ones at day 15 in the sixth individual. The score performed with similar accuracy as on published epitope-specific clonotypes (Fig. 5C).

\subsection*{Retrospective detection of YFV 17D-reactive TCRs using post-vaccination data}
So far we have identified responding TCRs as those that significantly expanded between days 0 and 15. While pre-challenge timepoints are easy to collect in vaccination studies, this is not the case for  acute infections, where the first samples can usually be obtained only after the onset of symptoms, when it is too late to detect TCR expansion.
However, the clonal \textit{contraction} dynamics (Fig.~2, day 15 to day 45) can in principle be used to identify responding clonotypes, by comparing a timepoint taken on the peak of the response to a timepoint taken several weeks or months after the infection.

To demonstrate the feasibility of this detection on our data, we identified significantly ($p<0.05$, fold change $>1$, see Methods) contracted clonotypes between day 15 and day 45 using our model. We computed the overlap between this set of candidate TCR$\beta$ clonotypes with the subset of expanded clonotypes obtained before (Fig. 6). Strikingly, 73\% to 97\% of the significantly contracted TCRs were also present in the expanded subset, showing that the contraction dynamics can help to identify YFV-reactive clonotypes with high specificity. This method is also sensitive: 47\% to 80\% of expanded TCRs could be identified by contraction. This shows that contraction dynamics alone is sufficient to identify a large fraction of the responding TCRs.  Thus, our method could be used to identify clonotypes responding to infections in the clinic, when pre-infection timepoints are not available.

 \section*{Discussion}
In this study we used high-throughput TCR$\beta$ repertoire profiling to identify major changes occurring in the repertoire after immunization with the live attenuated YFV-17D vaccine. We found several hundreds of unique TCR clones in each donor expanded in response to vaccination. A strong clonal expansion of up to $>2000$ fold occurred between days 7 and 15 following vaccination. This proliferation corresponds to at least 11 divisions in 7 days, with an average of 15 hours per cell cycle. Similar division rates (doubling times of 8-15 and 11-17 hours for CD4+ and CD8+ T-cells respectively) were observed for LCMV infection \cite{DeBoer2003} and also in adoptive transfer experiments in mice (18 hours) \cite{Buchholz2013}. However, in our case the actual expansion rate could be much higher, because when a clonotype is not found on day 7, its initial concentration is unknown, providing only a lower bound estimate of the fold change.
{\color{black} Higher sequencing depth or shorter intervals between timepoints are needed to refine this estimate of expansion rates. This extension would allow us to study the impact of the T-cell clone phenotype and the TCR sequence on  the clonal expansion rate. Initial low concentrations of YFV-reactive clonotypes suggest their naive phenotype.}

\AW{Our longitudinal approach can be applied to identify antigen-specific clonotypes for poorly characterized pathogens with unknown epitopes, which may be useful to study the immune response to emerging viral infections. One could also use this approach to track the response to infection and vaccination of unconventional T-cell subsets for which antigens are still unknown (i.e. $\gamma\delta$ T-cells).

While analysing expansion between a pre-vaccination timepoint and the peak of the T-cell response (day 15 in our case) is the the most natural choice, for real-world, non-experimental acute infections, acquiring a pre-infection time point is often impossible or impractical. In this case, we showed how to use the dynamics of clonal contraction to identify reactive clonotypes, by comparing their abundance at the peak of the response to a timepoint taken a few weeks later. Implementing such a protocol in the clinic would lead to an exponential growth of the number of annotated TCR sequences specific to a range of infectious diseases, facilitating diagnostics and vaccine design.}

Simultaneous sequencing of bulk PBMC, CD4+ and CD8+ subsets allowed us to determine the phenotype of the responding clones and to describe the kinetics of the response inside each compartment. While Blom et al. \cite{Blom2013} found that YFV-specific CD4+ T-cells concentration peaked slightly earlier than CD8+ T-cells, we did not detect any difference in the expansion kinetics with our limited temporal resolution. More timepoints between days 7 and 15 could have been helped to detect such differences. However, we did find differences in the contraction kinetics, with CD4+ cells contracting faster than CD8+ cells. We also found differential recruitment to the CD45RO+ memory compartment, with many more CD4+ than CD8+ TCRs detected in the CD45RO+ fraction on day 45. CD8+ memory formed in response to YFV-immunization was shown to be largely CD45RA+ \cite{Akondy2009, FuertesMarraco2015, Blom2013}, whereas James et al. \cite{James2013} found that CD4+ memory cells were mainly CD45RA-. Our results suggest that CD4+ and CD8+ YFV-reactive memory populations also differ in their CD45RO expression.

To the best of our knowledge, our study is the first to describe the reaction of the T-cell repertoire {\color{black} in a model of }an acute infection in genetically identical individuals. 
It was previously shown that the T-cell repertoires of twins have more sequence overlap in abundant clonotypes \cite{Zvyagin2014,Rubelt2016}. Part of this extensive overlap may be explained by \textit{in utero} sharing of T-cells \cite{Pogorelyy2017}. Authors in \cite{Qi2016} also found more overlap in vaccinia virus-specific CD4+ repertoires (isolated by \textit{in vitro} cultivation with the antigen) in twins.
Consistent with these findings, we report more TCR$\beta$ amino acid sequence overlap in the YFV-reactive repertoires of twins than in those of unrelated individuals. Twin YFV-specific TCR$\beta$ clonotypes also show higher nucleotide sequence overlap than those of unrelated individuals, even relative to their higher amino acid sequence overlap (see SI Table S4).
To assess the significance of this observation, we used a generative model to ask how likely it is to produce such an amount of shared nucleotide sequences by convergent recombination (see Methods). In the vast majority of simulations ($p<0.021$) the model underestimates the number of shared sequences between twin donors. We speculate that some of these shared TCRs were exchanged \textit{in utero}. Yet two thirds of YF-reactive TCR$\beta$ amino acid sequences shared between twins have different nucleotide sequences and can only be explained by convergent recombination and selection. Our results also suggest a mechanism for the previously reported extensive sharing of abundant TCRs in twins \cite{Zvyagin2014}. Under this mechanism, twins would share more TCR sequences expanded in response to the same infection. 

We showed that the response to the vaccine is very diverse, with few TCRs shared even between donors with identical genetic and environmental backgrounds. Nevertheless, using a simple similarity measure, it is possible to identify YFV-specific clonotypes from yet unseen repertoires with high specificity, using datasets of TCRs with known YFV specificity. The sensitivity of this classifier could be improved by collecting more examples of YFV-specific TCRs from more donors. We also show how published antigen-specific sequences can be used for functional repertoire annotation. On day 15 after immunization we found a much higher cumulative frequency of published A02-NS4b$_{214-222}$-specific sequences compared to pre-vaccination levels. Yet only a few significantly expanded clones matched those published sequences. This could be explained in two possible ways. First, our significantly expanded clones may be specific to epitopes other than NS4b$_{214-222}$. 
Second, the A02-NS4b$_{214-222}$ specific T-cell repertoire may be so diverse that little overlap is to be expected between random subsamples of it. Further accumulation of antigen-specific TCR sequence data --- acquired by sequencing of multimer-specific cells, longitudinal studies as done here, and from disease-associated studies with large cohorts \cite{Emerson2017, Faham, Pogorelyy2017b} --- will provide the means for disease diagnostics and extraction of clinically relevant information from T-cell repertoire data.  

 \section*{Materials and methods}
\subsection*{Blood donors and samples}
Three pairs of healthy monozygous twins participated in this study. Monozygosity was confirmed using polymorphic Alu-insertion genotyping \cite{Mamedov2010}. All donors were HLA-typed using an in-house HTS-based solution (see Table S1). None of the participants have ever been vaccinated with YFV-vaccine or traveled to Africa or South America. The blood was collected with informed consent in a certified diagnostics laboratory. For each unsorted PBMC sample we used 4 ml of whole blood per biological replicate, and another 4 ml for CD4+ and CD8+ subpopulation isolation using Dynabeads CD4+ and CD8+ positive selection kits (Invitrogen), and 4 ml for CD45RO+ subpopulation isolation using the CD45RO+ positive isolation kit (Miltenyi Biotec). PBMC were isolated with the Ficoll-Paque method according to the manufacturer's protocol. All isolated cells were immediately lysed with the Trizol reagent (Invitrogen).

\subsection*{MHC-dextramer staining}
Staining with HLA-A*02 the dextramer loaded with the NS4b$_{214-222}$ peptide (LLWNGPMAV) from YFV-17D (Immudex) was performed according to the manufacturer's protocol. 25 ml of whole blood was used for PBMC isolation using the Ficoll protocol, PBMCs were stained with MHC-dextramer-PE, and also anti-CD8-FITC (Beckman Coulter, dilution 1:100) and anti-CD3-PC5 antibodies (eBioscience, dilution 1:100). 

\subsection*{HLA-DR staining}
Isolated PBMCs were stained with anti-CD8-FITC (Invitrogen, dilution 1:100), anti-CD38-PE (eBioScience, dilution 1:100), anti-HLA-DR-PerCP (Invitrogen dilution, 1:100) according to the manufacturer's instructions.

\subsection*{IFN-gamma secretion assay }
 
The IFN-gamma secretion assay was performed using the IFN-gamma Secretion Assay - Cell Enrichment and Detection Kit (PE human, Myltenyi Biotec). 2.5 ml of whole blood was incubated with $10^4$/ml YFV-17D particles and 5 ug/ml of the anti-CD28 antibody (eBioscience)  for 5 hours. The same amount of blood was incubated with the same amount of media without virus particles and with 5 ug/ml of the anti-CD28 antibody and used as negative control. After stimulation erythrocytes were lysed with ACK Lysing Buffer (Gibco, Thermo Fisher Scientific), and white blood cells were washed by PBS. The secretion assay and IFN-gamma positive cells isolation were performed according to the manufacturer's protocol using MACS LS Columns (Myltenyi Biotec). anti-CD8-PС5 (Beckman Coulter, dilution 1:100) and anti-CD3-FITC (eBioscience, dilution 1:100) antibodies were added along with the IFN-gamma detection antibody at the corresponding step. Aliquot (1/10) of isolated cells were analyzed by FACS and all other cells were lysed in 500 ul of Trizol reagent (Invitrogen) and used for subsequent RNA isolation.

\subsection*{Analysis of sequencing data of functional assays}
All three functional assays (IFN-gamma secretion assay, CD8+CD38+HLA-DR+ staining and A02-NS4b$_{214-222}$ dextramer staining) provided us with TCR$\beta$ repertoire sequencing information. However, both fluorescence activated cell sorting after dextramer staining, activation marker staining and magnetic column enrichment are not completely precise. These methods enrich resulting samples with YF-specific cells, but abundant irrelevant cells are still found in the corresponding repertoires. To detect clonotypes enriched in the CD8+CD38+HLA-DR+ population, the IFN-gamma producers population or the tetramer-positive population in comparison to the bulk population at  the same timepoint we used  Fisher's exact test, as suggested in Ref.~\cite{DeWitt2015}. {For each clone the number of barcodes occupied by it in the activated population, the number of barcodes occupied by other clonotypes in the activated population, the number of barcodes occupied by the clone in the bulk population and the number of barcodes occupied by all other clones in the bulk population formed a $2\times2$ contingency table, and then one way Fisher's exact test was applied.} This procedure was repeated for each clone found in the activated population, clones with BH-corrected p-values$<$0.05 were considered significantly enriched in the activated population.

\subsection*{TCR$\beta$ cDNA library preparation and sequencing}
TCR$\beta$ cDNA library preparation was performed as described in Ref.~\cite{Pogorelyy2017}. Briefly, PBMC were isolated from whole blood using the Ficoll density gradient, the total RNA was isolated using the Trizol reagent (Invitrogen). cDNA synthesis with template switch was followed by two steps of PCR amplification to introduce adapter sequences necessary for sequencing on the Illumina platform. Unique molecular identifiers (UMI) were introduced into each cDNA molecule using the template switch effect on the cDNA synthesis step. This allowed us to know the number of cDNA molecules corresponding to each clone, instead of number of raw sequencing reads and thus to correct for PCR bias.

Samples were sequenced on the Illumina HiSeq 2500 with $2\times100$ read length. Samples from twin donors were processed and sequenced separately, to prevent contamination and sample barcode exchange on Illumina flow cell. 

\subsection*{Raw data analysis}
Raw data was demultiplexed and clustered by UMI using the MIGEC software \cite{Shugay2014}, alignment of V,D,J templates was then performed by MiXCR \cite{Bolotin2015}.

\subsection*{Statistical analysis}
Candidate responding clones were identified using a model of mRNA count statistics accounting for differential expression and the sequencing process \AW{[Puelma Touzel et al, in preparation]}. The normalized clone size, $f$, is distributed according to the probability density function $\rho(f)$, bounded by $M_T^{-1}\leq f \leq 1$, where $M_T=10^{11}$ is an estimate of the total number of lymphocytes in an individual. Based on previous observations \cite{Weinstein2009,Mora2010,Mora2016}, $\rho(f)$ is set as a power-law, i.e. $\rho(f)\propto f^{-\gamma}$. A clone of size $f$ appears in a sample containing $M$ lymphocytes on average as $fM$ cells. To account for overdispersed count statistics, the number of cells is set to be Negative-Binomial distributed with mean $fM$ and variance $fM+a (fM)^{\beta}$, with $a>0$ the coefficient and $\beta>1$ the power controlling the over-dispersion. For each clone, the number $n$ of detected mRNA molecules (i.e. UMI) is distributed according to a Poisson distribution with mean $mN/M$, where $N/M$ is the average number of UMI per cell, obtained using the observed total number of molecules, $N$.
 
We inferred the parameters of this model, $\theta=(\gamma,M,a,\beta)$, from day-0 replicates by maximizing the likelihood of the observed count pairs, $(n_1,n_2)$, where $n_i\sim \mathrm{Poisson}(m_i N_i/M)$, $m_i\sim \mathrm{NegBin}(fM,fM+a (fM)^{\beta})$, for each replicate $i=1,2$, and $f\sim\rho$ is common to both replicates. For a given pair, the likelihood, $P(n_1,n_2|\theta)$, is obtained by marginalizing over $m_1,m_2$, and $f$.
 
We introduce a selection factor $s$ defined as the log-fold change between a clone's frequency on one day, $f$, and that on another, $fe^s$, and define $P(n_1,n_2|s,\theta)$ as before, but replacing $f$ by $fe^s$ in the definition of $m_2$. Given a prior distribution  $P(s|\theta')$ over $s$ parametrized by a set of parameters $\theta'$ distinct from $\theta$, we used Bayes rule to obtain the posterior log fold-change probability function given an observed count pair, $P(s|n_1,n_2,\theta,\theta')$. In analogy with $p$-values, we used the posterior probability corresponding to the null hypothesis that they are not expanded, $p=P(s\leq 0|n_1,n_2,\theta,\theta')$ to rank the clones by the significance of their expansion, using a threshold of $p<0.025$. We parametrize our prior as $P(s|\theta')=(\alpha/2\bar s)\exp(-|s|/\bar{s})+(1-\alpha)\delta(s)$, with $0\leq \alpha \leq 1$ the fraction of clones that respond to the change, and $\bar{s}>0$ their typical effect size. We set the values of the parameters $\theta'=(\alpha,\bar s)$ of this prior by again maximizing the likelihood of the count pair data given the model over $\theta'$, $\int P(n_1,n_2|s,\theta)P(s|\theta')\textrm{d}s$.

\subsection*{Identification of expanded clones using edgeR}
To check robustness of our clonal expansion model we also used an independent approach to detect  expanded clones. edgeR \cite{Robinson2009} is a package used to analyse a variety of count data produced with HTS. It was applied for differential gene expression analysis, differential splicing analysis and bisulfite sequencing. Here we apply statistical tests implemented in the edgeR package to identify clones expanded after YFV immunization. Two biological replicates were used for each timepoint, TMM-normalization and trended dispersion estimates was performed as described in the edgeR manual. An exact test based on the quantile-adjusted conditional maximum likelihood (qCML) was used to identify clones significantly expanded between pairs of timepoints. A clonotype was considered significantly expanded, if it's $\rm{log}_2$ fold change estimate $\rm{log}_2$FC$>$5 and it's p-value after multiple testing correction was lower than 0.01. We reproduced the main results described in the manuscript text using clonotypes identified as significant by edgeR (see SI Fig. S6), so two differential clone expansion models agree with each other. 

\subsection*{Sequence similarity analysis}
We used the Hamming distance to quantify CDR3 sequence similarity in each YFV-reactive repertoire. We counted the number of CDR3 amino acid sequence pairs, which are 0 mismatches (same amino acid, but different nucleotide sequences), 1 mismatch or less, 2 mismatches or less from each other. Since the number of YFV-reactive clones varied among donors, the number of similar clones inside each YFV-reactive repertoire needed to be normalized. We divided it by $n(n-1)/2$, which is the total number of possible clone pairs. We randomly sampled clonotypes from the same timepoint for comparison, and found a smaller number of similar pairs. In Fig. 4A. we plot the normalized sharing in log scale to compare these values. 

\subsection*{Nucleotide sequence sharing analysis of YFV-reactive clonotypes}
To see, if the excess of nucleotide sequence sharing in twins may be explained by convergent recombination, we performed the  following simulation: for each donor, for each shared YFV-specific TCR$\beta$ amino acid sequence we generated nucleotide variants using a previously published TCR recombination model \cite{Murugan2012} and compared the resulting simulated nucleotide sequence variants between donors. Then the total number of shared nucleotide sequences between twins in this simulation was compared to the total number of shared nucleotide sequences between the twins in the data. The simulation was performed 1000 times: in 979 cases out of 1000 the number of nucleotide sequences shared between twins was less than in the data, which gives a p-value$=$0.021. This result suggests, that at the least some of the nucleotide sequences of YF-reactive clones may be shared prenatally between twin donors, because this extensive sharing of nucleotide sequences cannot be explained by convergent recombination and selection. On the contrary, the number of shared nucleotide sequences between unrelated donors in the simulation was rarely less than in the data (in 295 cases out of 1000), which means that the nucleotide sequence sharing between unrelated individuals can easily be achieved by convergent recombination alone (p-value=0.7). 

\section*{Acknowledgments}
TCR$\beta$ libraries sequencing, raw sequencing data processing and reconstruction of TCR$\beta$ repertoires were supported by the Russian Science Foundation grant n.15-15-00178.  M.V. Pogorelyy and E.S. Egorov are supported by Skoltech Systems biology fellowships. This work was partially supported by the European Research Council Consolidator Grant n. 724208. We  thank Dr. I.V. Zvyagin for his help in finding appropriate donors for this study and M.F. Vorovitch for YFV-17D purification. The experiments were in part conducted using the equipment provided by the CKP IBCH RAS core facility.

\bibliographystyle{pnas}

\setcounter{figure}{0}
\setcounter{table}{0}
\renewcommand{\thefigure}{S\arabic{figure}}
\renewcommand{\thetable}{S\arabic{table}}

\section*{ }

\begin{figure}[p]
\noindent\includegraphics[width=\linewidth]{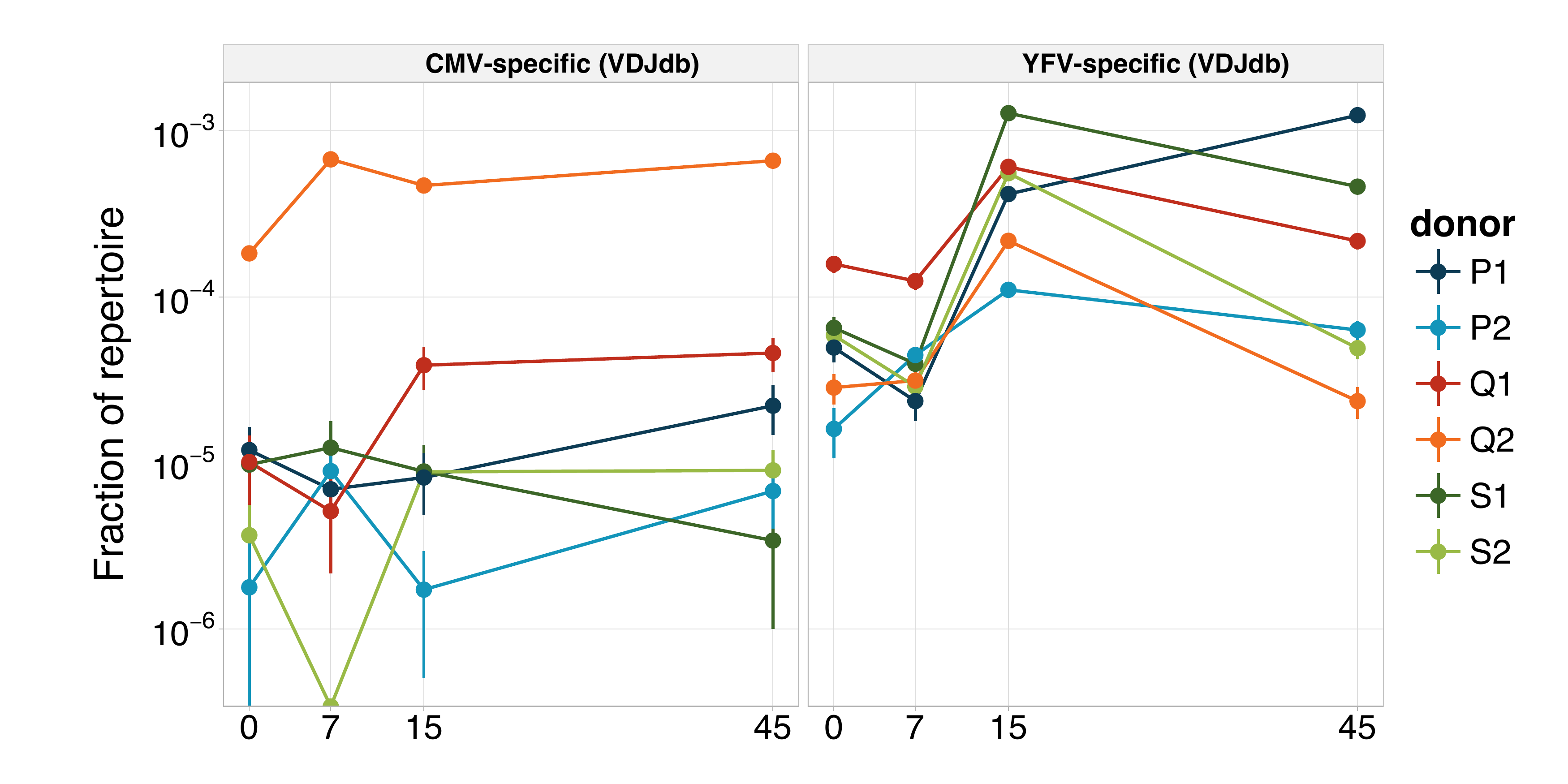}
\caption{{\bf ${\rm Log}_{10}$ cumulative frequency of published YFV- and CMV-specific clonotypes in our CD8+ datasets.} We found much higher relative abundance of known YFV-reactive clonotypes on day 15 comparing to day 0 for each donor, however no such effect was found for known CMV-reactive clonotypes.}
\label{figS2}
\end{figure}

\begin{figure*}[p]
\noindent\includegraphics[width=0.8\linewidth]{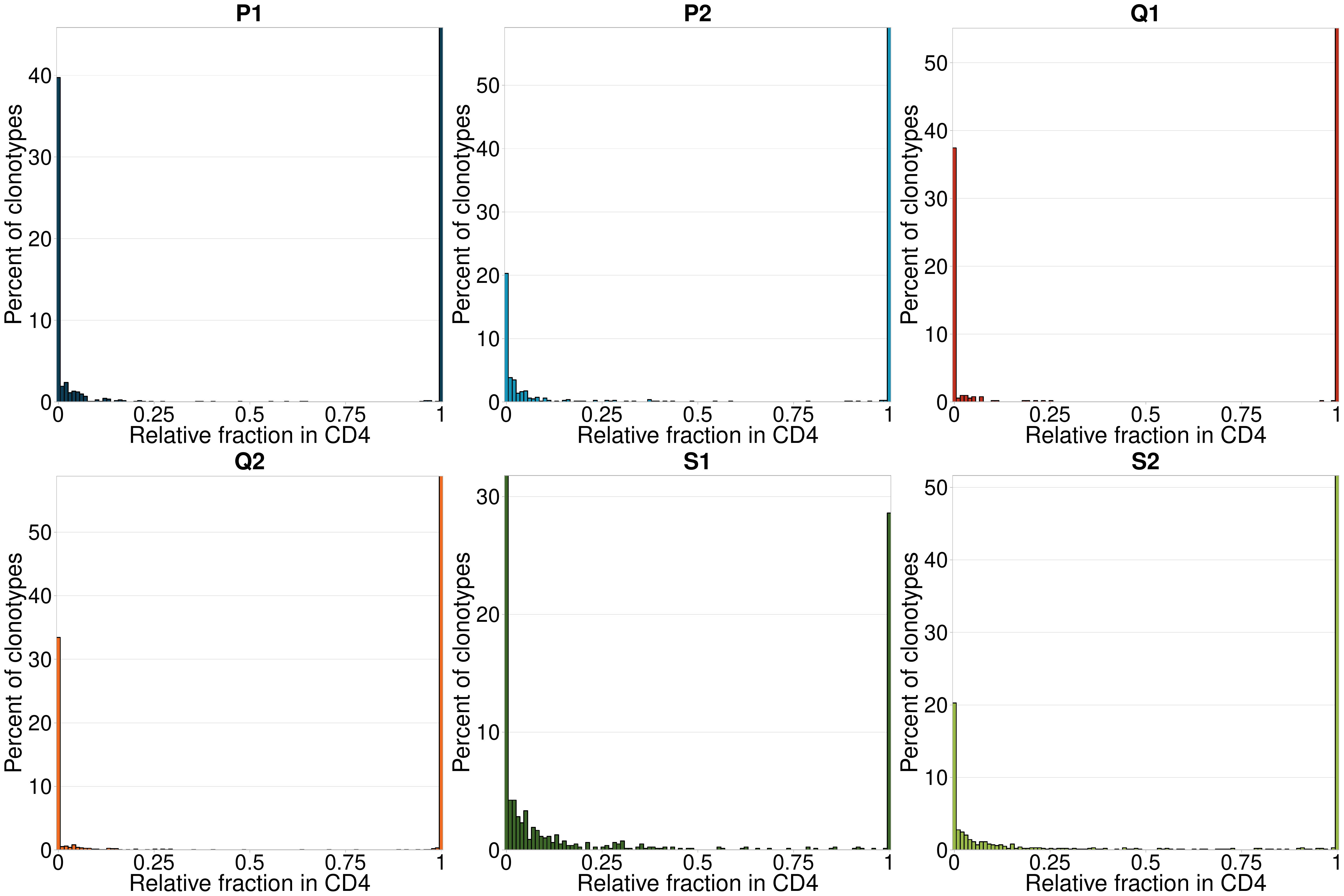}
\caption{{\bf Bimodality of YF-specific clone concentrations in the CD4+ and CD8+ compartments.} For each expanded clonotype from bulk PBMC we show relative fraction in CD4+  repertoire: fraction of 0 means that clonotype is found exclusively in CD8+ repertoire, fraction of 1 means clonotype is found exclusively in CD4+ repertoire.}
\label{figS3}
\end{figure*}

\begin{figure}[p]
\noindent\includegraphics[width=\linewidth]{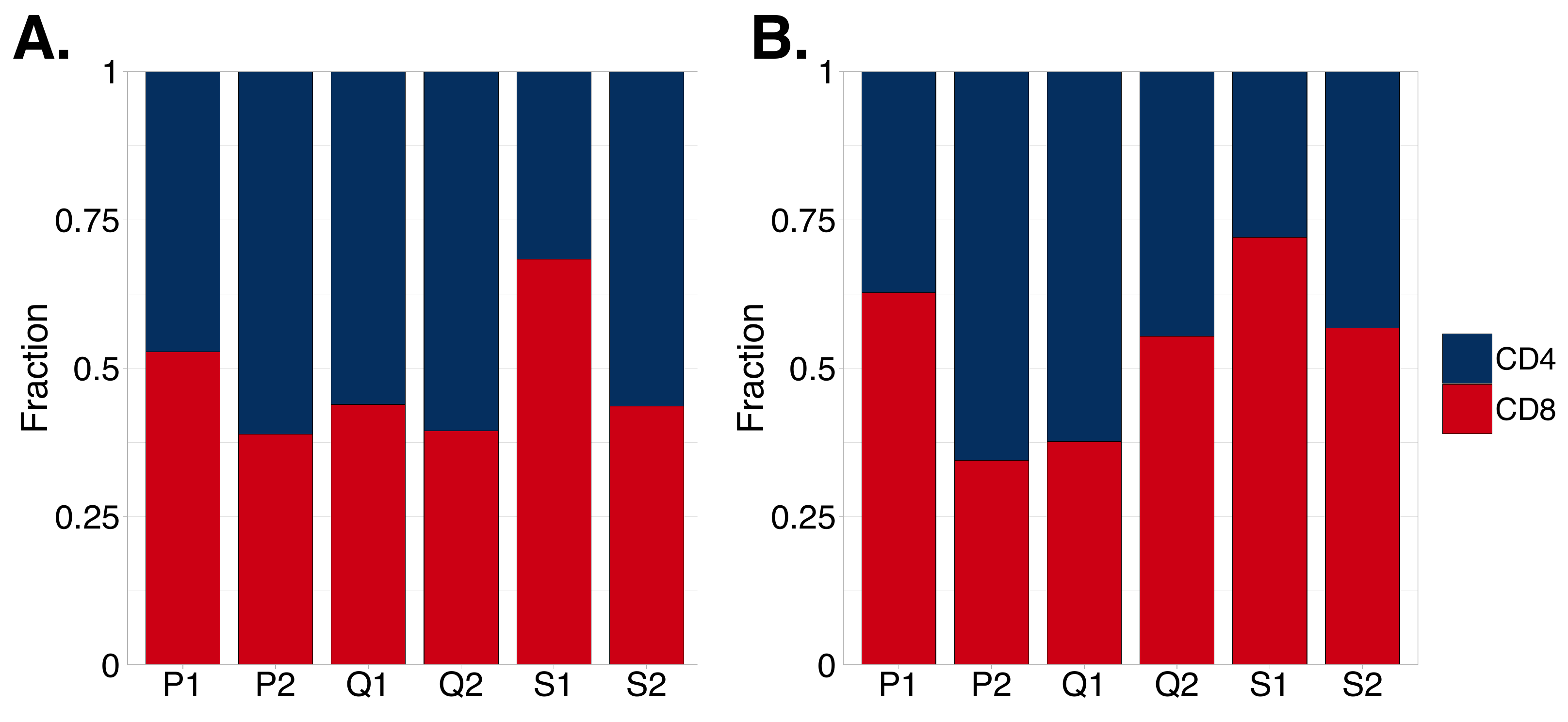}
\caption{{\bf Distribution of yellow fever reactive clonotypes between CD4+ and CD8+ subpopulations.} 
{\bf A.} Relative fractions of activated clonotypes attributed to the CD4+ and CD8+ compartments  and {\bf B.} relative fractions of cells attributed to the CD4+ and CD8+ activated clonotypes in each repertoire.  
}
\label{figS4}
\end{figure}

\begin{figure}[p]
\noindent\includegraphics[width=\linewidth]{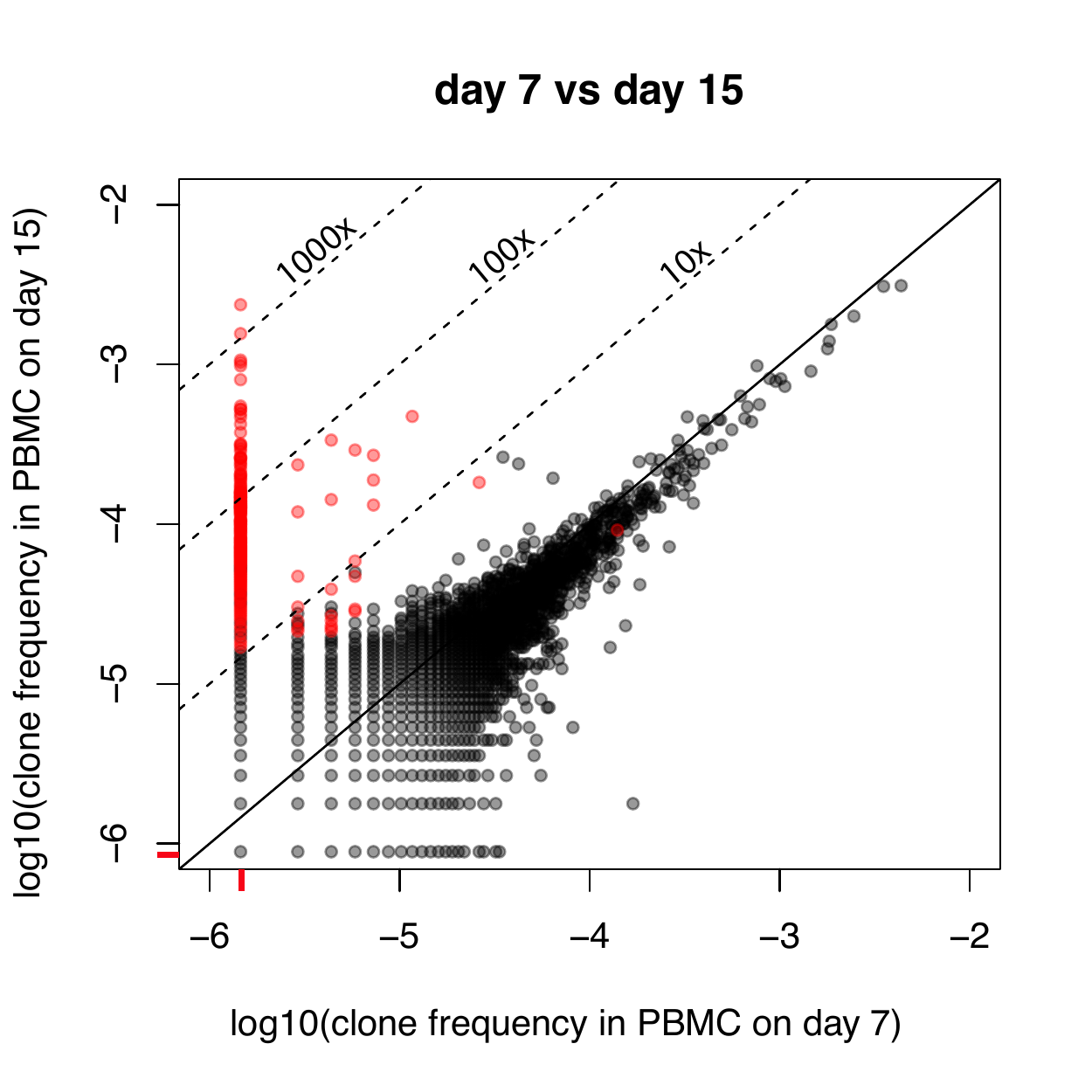}
\caption{{\bf Scatterplot of clone frequencies in donor S1 on day 7 and day 15.} ${\rm Log}_{10}$ of clone frequency (defined as the  ratio of the UMI for given clone and the sum of UMIs in sample) on day 7 (x-axis) is plotted against ${\rm log}_{10}$ of clone frequency on day 15. Pseudocounts of 1 UMI are used to show 0 on logarithmic scale. Red ticks on axes mark pseudocount size, these clones have 0 concentration at the corresponding timepoint. Red dots on the plot indicate clones significantly expanded between day 0 and day 15.
}
\label{figS5}
\end{figure}

\begin{figure}[p]
\noindent\includegraphics[width=0.5\linewidth]{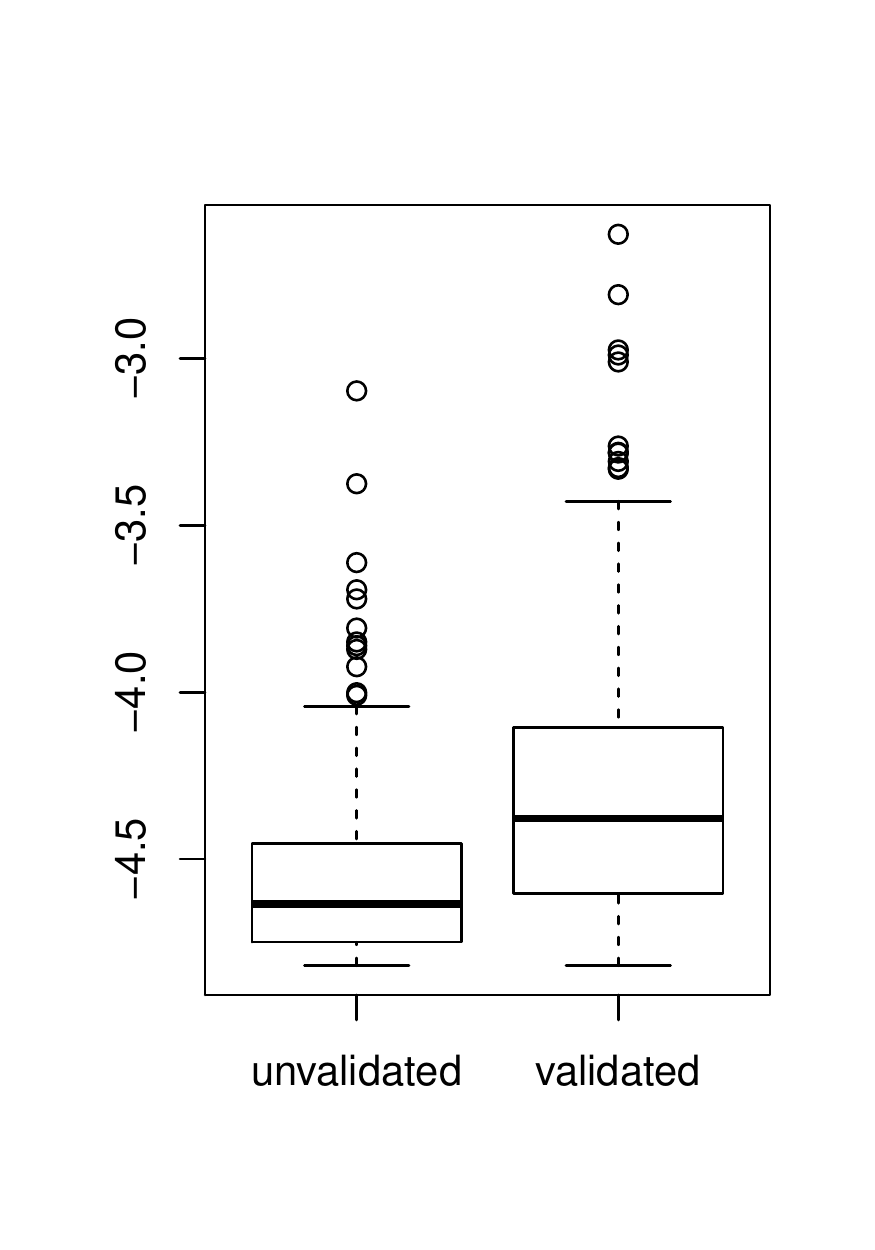}
\caption{{\bf ${\rm Log}_{10}$ frequencies of YFV-reactive clones unvalidated and validated by functional tests.} Unvalidated clones have significantly lower frequencies (t-test p$<10^{-16}$)}.
\label{figS6}
\end{figure}

\begin{figure}[p]
\noindent\includegraphics[width=\linewidth]{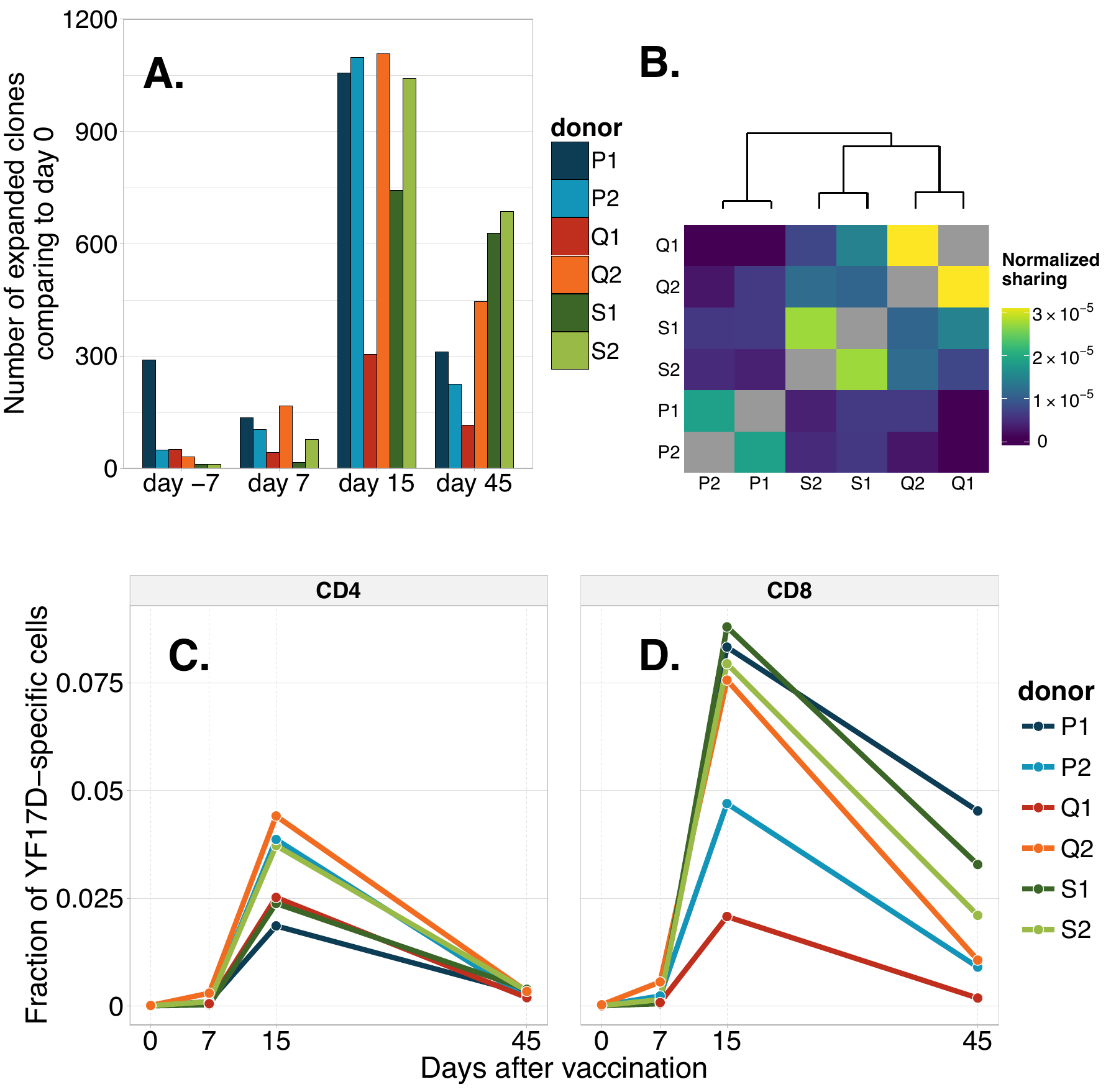}
\caption{{\bf Results of our analysis could be reproduced using the edgeR differential gene expression analysis software. A.} Number of significantly expanded clonotypes in comparison to day 0. {\bf B.} The heatmap shows the number of the same amino acid clonotypes between two sets of clones significantly expanded from day 0 to day 15, divided by the product of the sizes of these sets. {\bf C. and D.} Dynamics of T-cells in the CD4+ and CD8+ compartments.  Total fractions of CD4+ ({\bf C.}) and CD8+ ({\bf D.}) repertoires occupied by clonotypes significantly expanded from day 0 to day 15 is plotted at different timepoints. }
\label{figS7}
\end{figure}

\begin{table}[p]
\begin{tabular}{cccc}
\hline
Locus & S1 and S2 & P1 and P2 & Q1 and Q2 \\ \hline
A & 02:01:01/03:01:01 & 02:01:01/02:01:01 & 02:01:01/02:01:01 \\
B & 38:01:01/07:02:01 & 51:01:01/07:02:01 & 13:02:01/44:02:01 \\ 
C & 07:02:01/12:03:01 & 07:02:01/14:02:01 & 05:01:01/06:02:01 \\ 
DQB1 & 06:03:01/06:02:01 & 03:01:01/04:02:01 & 02:02:01/06:03:01 \\ 
DRB1 & 13:01:01/15:01:01 & 08:03:02/08:01:03 & 07:01:01/13:01:01 \\ 
DRB3 & 01:01:02 & - & 01:01:02 \\ 
DRB4 & - & - & 01:03 \\ \hline
\end{tabular}
\caption{HLA-haplotypes of donors}
\end{table}

\begin{table}[p] 
\begin{tabular}{lllllll}
\hline
 & S2 & S1 & P2 & P1 & Q2 & Q1 \\ \hline
CMV, 0 mism & 0 & 0 & 0 & 0 & 0 & 0 \\ 
YFV, 0 mism & 6 & 3 & 1 & 2 & 4 & 3 \\ \hline
CMV, 1 mism & 0 & 0 & 1 & 0 & 2 & 0 \\ 
YFV, 1 mism & 9 & 24 & 9 & 12 & 21 & 5 \\ \hline
CMV, 2 mism & 8 & 5 & 3 & 5 & 10 & 2 \\ 
YFV, 2 mism & 29 & 27 & 11 & 24 & 37 & 19\\ \hline
\end{tabular}
\caption{The number of YFV-reactive TCR$\beta$ sequences found in this study that are similar to published TCR$\beta$ sequences specific for immunodominant YFV and CMV epitopes.}
\end{table}

\begin{table}[p] 
\centering
\begin{tabular}{lrrrrrrr}
  \hline
 & S2 & S1 & P2 & P1 & Q2 & Q1 \\ 
  \hline
Fraction of public YFV-reactive clones & 0.03 & 0.04 & 0.03 & 0.03 & 0.02 & 0.04 \\ 
Cumulative freq. of public YFV-reactive clones & 0.05 & 0.08 & 0.06 & 0.03 & 0.03 & 0.34 \\ 
   \hline
\end{tabular}
\caption{Contribution of public clones to YFV response. A TCR$\beta$ amino acid clonotype is considered public if it is shared with at least one other donor in this study.}
\end{table}

\begin{table}[p]
\centering
\begin{tabular}{rllllll}
  \hline
 & S2 & S1 & P2 & P1 & Q2 & Q1 \\ 
  \hline
S2 & 970/955 & 5/20 & 0/3 & 0/4 & 2/9 & 1/3 \\ 
  S1 & 5/20 & 774/763 & 0/3 & 0/5 & 1/12 & 0/4 \\ 
  P2 & 0/3 & 0/3 & 800/792 & 1/18 & 0/2 & 0/0 \\ 
  P1 & 0/4 & 0/5 & 1/18 & 1151/1142 & 0/7 & 0/1 \\ 
  Q2 & 2/9 & 1/12 & 0/2 & 0/7 & 1537/1503 & 4/20 \\ 
  Q1 & 1/3 & 0/4 & 0/0 & 0/1 & 4/20 & 520/510 \\ 
   \hline
\end{tabular}
\caption{The number of unique shared nucleotide/amino acid sequences between YFV-reactive repertoires of our donors. The diagonal shows the total number of unique nucleotide/amino acid TCR$\beta$ sequences inside each YFV-reactive repertoire.}

\end{table}

\end{document}